\documentclass[draftcls,onecolumn,12pt]{IEEEtran}
\ifCLASSINFOpdf
\else
\fi
\usepackage{cite}
\usepackage{amsmath,amssymb,amsfonts}
\usepackage{algorithmic}
\usepackage{graphicx}
\usepackage{textcomp}
\usepackage{xcolor}
\usepackage{subfigure}
\usepackage{caption}
\usepackage{ifpdf}
\usepackage{float}
\usepackage{booktabs}

\begin{document}
%
\title{Two New Approaches to Optical IRSs: Schemes and Comparative Analysis}
%
%
%

\author{Haibo Wang,~\IEEEmembership{Member,~IEEE,}
        Zaichen Zhang,~\IEEEmembership{Senior Member,~IEEE,}
        Bingcheng Zhu,~\IEEEmembership{Member,~IEEE,}
        Jian Dang,~\IEEEmembership{Member,~IEEE,}
        and Liang Wu,~\IEEEmembership{Member,~IEEE,}
       
\thanks{Haibo Wang, Zaichen Zhang, Bingcheng Zhu, Jian Dang, and Liang Wu are with National Mobile Communications Research Laboratory, Southeast University, Nanjing 210096, China.
Zaichen Zhang is the corresponding author.}
\thanks{This work is supported by NSFC projects (61960206005, 61971136, 61803211, and 61871111), Jiangsu NSF project (BK20191261), Zhejiang Lab (No. 2019LC0AB02), the Fundamental Research Funds for the Central Universities, and Research Fund of National Mobile Communications Research Laboratory, Southeast University.}}

\maketitle

\begin{abstract}
Oriented to the point-to-multipoint free space optical communication (FSO) scenarios, this paper analyzes the micro-mirror array and phased array-type optical intelligent reflecting surface (OIRS) in terms of control mode, power efficiency, and beam splitting. We build the physical models of the two types of OIRSs. Based on the models, the closed form solution of OIRSs' output power density distribution and power efficiency, along with their control algorithms have been derived. Then we propose the algorithms of beam splitting and multi-beam power allocation for two types of OIRSs. The channel fading in FSO system and the comparison of two types of OIRSs in actual systems are discussed according to the analytical results. Experiments and simulations are both presented to verify the feasibility of models and algorithms. 
\end{abstract}


%
\IEEEpeerreviewmaketitle

\section{Introduction}
\IEEEPARstart {A}{s} a new type of communication device, the intelligent reflecting surface (IRS) breaks through the traditional system architecture with transceivers, which puts part of the information processing and beam control functions into the channel. It has a wide range of application scenarios and arises extensive attention. The main structure of IRSs in the microwave band is a phased array composed of multiple phased units, which can be applied in beam shaping and signal modulation \cite{Yifei2020Intelligent,9235486,9117136,9119122,Christos2015Enhanced,9200674}. Different from that in microwave band, the optical intelligent reflective surfaces (OIRS) can be modeled as a single mirror for optical communication scenarios. Due to the high directivity of optical signal, beam reflection and deflection can already be achieved with a single mirror \cite{9013840,Naja2019,Naja2020}, thereby relieving the burden of the direct path. However, the capabilities of OIRS are more than that. The OIRS with the array structure can achieve the functions of beam focusing, beam splitting, real-time multi-beam deflection, and multi-beam power allocation \cite{cva2019,zcao2019,9140329,Mcmanamon2004Modular,4373909,1540073}. By these functions, each OIRS can be utilized as a 'sub base station', responsible for beam control and power allocation in a certain area, thereby relieving the burden of the base station and expanding the application scenarios of optical communication. The system assisted by distributed OIRSs can concentrate more energy to the users while maintaining large coverage, which is expected to achieve higher communication rate. Furthermore, since each OIRS can independently control the beam's number, size and location in real time, the system will be more flexible to handle different situations. Space division multiple access (SDMA) for multiple regions can be implemented by sending signals with the same frequency to OIRSs in different locations, which further improves the spectrum utilization \cite{4133000,7828152}.\par
There are two main architecture of array-type OIRSs, which are micro-mirror array (MA) and optical phased array (OPA). The former is composed of multiple freely adjustable micro-mirrors. The advantages of low cost and larger coverage make its application scenarios broader. The latter is based on the programmable phased modulation array. Its architecture is similar to IRS in the microwave band, thus can also be used for signal modulation. However, the OPA's phased unit is tiny (must fit the light wavelength) and costly, thus it is difficult for the OPA-type OIRS to occupy large area. Apart from the above differences, the control methods, the light field of their output beam, and the performance of communication system assisted by them vary a lot, which will be discussed in this work.\par 
The single mirror-type OIRS was systematically modeled and analyzed in \cite{9013840}. The channel fading and system performance of the OIRS-assisted free space optical communication (FSO) system were derived. However, since the single-mirror OIRS output beam is still Gaussian, its performance analysis is not applicable to the array-type OIRS. In \cite{9276478}, the author compares the the mirror array-type OIRS with the metasurface type OIRS in the visible light communication (VLC) system. OIRSs are utilized to replace the partially diffuse reflection surface in VLC scenario to adjust the radiance on the receiving plane. Since this research is based on the VLC system with LED sources, its modeling and derivation for OIRS are completely different from that in FSO systems. In addition, the work does not use OIRSs for beam splitting and power allocation. When controlling the metasurface-type OIRS, it also does not consider holographic control and Fraunhofer diffraction. In \cite{6358647,5069618,7320951,8354597}, the holographic control method and diffraction pattern of the optical phased array have been analyzed and experimentally verified, which is used to realize holographic imaging and beam scanning. However, since the OPAs in these work are not applied to communications, the output power efficiency, power density distribution, beam splitting and power allocation that are concerned in optical communications are not analyzed. There is no systematic comparison between optical phased array and micro-mirror array, either.\par
Although the research on MA and OPA has been mature and their products have been put into operations \cite{8925046,7288883,8925234,8678389,9222022}, there is no systematic comparison for two types of array OIRSs in the FSO system. The output power distribution and efficiency of both OIRSs, along with their control algorithms, need to be specified. Furthermore, the physical modeling of OIRSs plays an important role in their auxiliary communication system, where the array-type OIRS can not be simply modeled as a single mirror. The performance analysis based on the impractical OIRS physical model will be inconsistent with the actual system. For example, the amplitude distribution of beam from both two types of OIRSs are not Gaussian, which brings in different channel fading models. Fraunhofer diffraction should be considered for the OPA-type OIRS, and the superposition of the output light field of multiple micro-mirror elements should be considered for the MA-type OIRS. Since the respective characteristics of the two types of OIRSs have not been systematically compared, there is a lack of clear guidance on which type of OIRS structure to choose for different optical communication systems. Based on the existing research, this paper further discusses the characteristics of the two types of OIRSs, including detailed physical modeling, output power distribution and efficiency,and algorithm for beam splitting and power allocation. Experiments with precise measurements are ultilized to verify the accuracy of the physical model and control algorithm. The innovations of this article are as follows:\par
\begin{itemize}
\item This paper proposes a point-to-multipoint free space optical communication (FSO) scenario based on array-type OIRSs. The scenario breaks through the traditional point-to-point FSO communication framework and makes full use of the functions of the array-type OIRSs, including beam focusing, beam splitting and multi-beam power allocation. Each OIRS can independently perform beam control and power distribution in the specific region, which expands the application of FSO communication to make it applicable to large areas and multiple users.\par
\item Based on the practical system, we have carried out a detailed modeling analysis of MA and OPA-type OIRS for point-to-multipoint FSO communication system. By taking into account a series of actual factors such as array element distribution, element size, element gap, input optical power distribution, receiver position, etc., we derive the control algorithms for two types of OIRSs along with their corresponding output power distribution and efficiency respectively. Experiments with precise measurements are done to verify the feasibility of the physical model and control algorithms. Experiments show that the deduced results in this paper are highly consistent with actual measurement results.\par
\item Based on the physical modeling analysis, we discussed the the channel fading caused by pointing error in communication system assisted by two types of OIRSs, which is different from traditional analysis and becomes a main problem for performance analysis.\par
\item We compare and analyze MA-type and OPA-type OIRSs from various aspects to provide reference for the selection of OIRSs in actual optical communication systems.
\end{itemize}
The rest of this paper are organized as follows. Section \ref{model} introduces the point-to-multipoint distributed FSO communication scenario and establish physical models of two types of OIRSs. In Section \ref{analysis}, we derive the control algorithms of two types of OIRSs and their corresponding output power efficiency and power density distribution. In Section \ref{split}, beam splitting and power allocation algorithms of two types of OIRSs are derived. Section \ref{discussion} discusses the small-scale channel fading in FSO system assisted by two types of OIRSs and the selection of two types of OIRSs in actual systems. Section \ref{simulation} introduces the experimental setup and shows some numerical results, and Section \ref{conclusion} draws conclusions.

\section{System model}\label{model}
\subsection{Point-to-multipoint distributed FSO communication scenario}
\begin{figure}[htbp]
\centering
\includegraphics[width=0.95\textwidth]{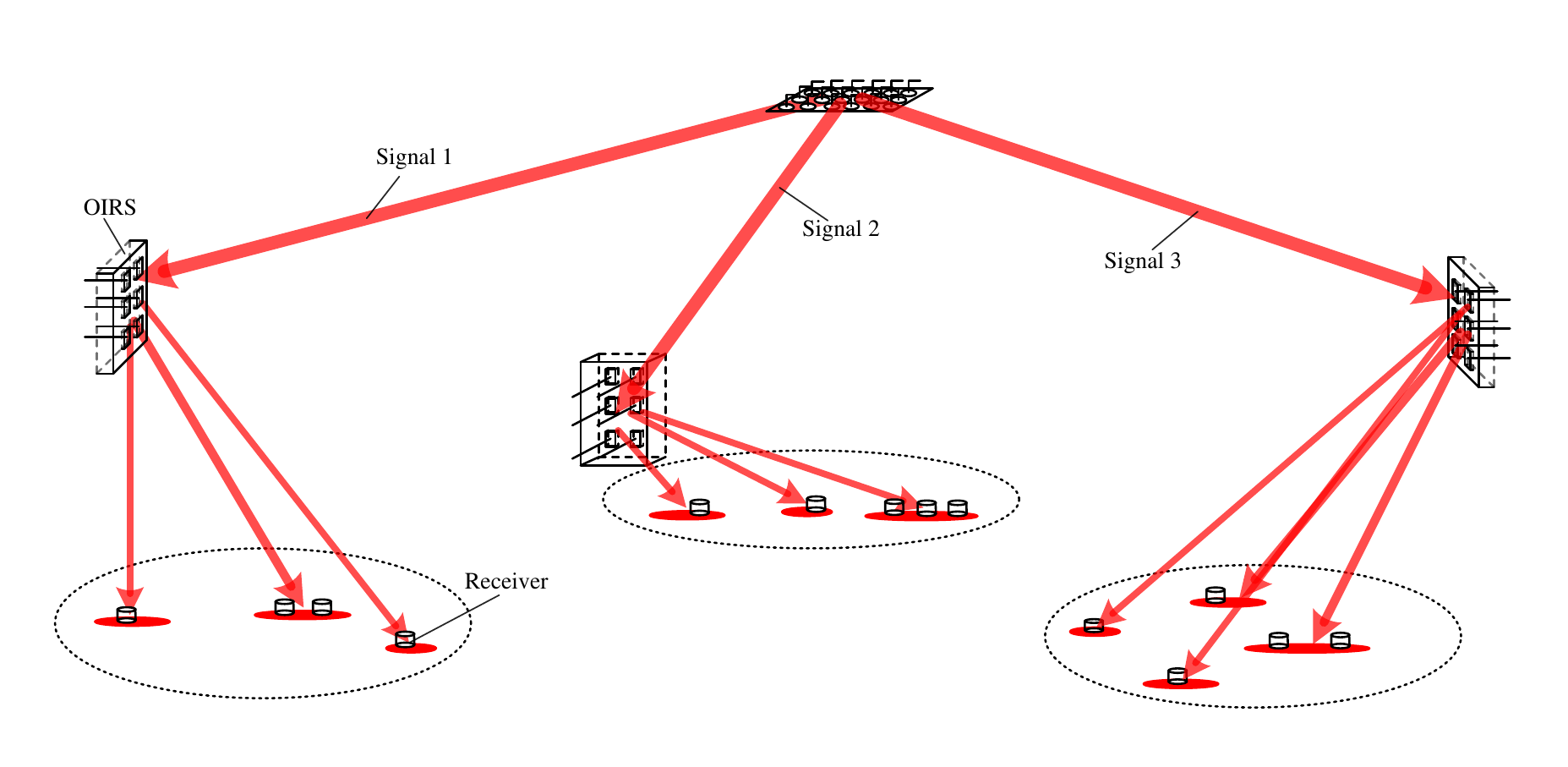}
\caption{Point-to-multipoint distributed FSO communication system.}
\label{fig.1}
\end{figure}

Based on the array-type OIRSs technology, we propose a point-to-multipoint distributed FSO scenario to expand the application of optical wireless communication. As shown in Fig. \ref{fig.1}, the base station utilizes a laser array as light source and simply establishes the connection with OIRSs. Each OIRS can concentrate the beam to the specific direction and split the beam into multiple sub-beams, whose size and power can be adjusted freely. Thus the OIRS can perform as a 'sub-base station' for beam control and power allocation in a specific sub-region. The advantages of this system are as follows:\par
\begin{itemize}
\item The base station only establishes connection with OIRSs. Since the position of OIRSs are fixed, the system distribute the base station's pressure on users' targeting and tracking to the OIRSs.\par
\item By sending different optical signals to OIRSs at different locations, SDMA can be realized for users in different sub-regions to share the same frequency band, which further improves the spectrum utilization.\par
\item Since the system utilizes laser as light source, the beam energy can concentrate to the specific direction. Thus the scenario has higher power efficiency with less light pollution compared with VLC system.\par
\item The array-type OIRS can achieve beam splitting with free adjustment of beam's number, size and power. Thus the system has high freedom degree and robustness with lower requirements for tracking accuracy.\par
\end{itemize}
In summary, the communication system makes full use of the array-type OIRS, which can function as a distributed 'sub-base station' to reduce the load of the base station. Compared with traditional FSO system \cite{6844864,1025501,4267802}, the system can support multiple users and has lower requirements for beam alignment's accuracy. Compared with the VLC system \cite{8703128,8930634,8818348,1277847,8698841}, its energy utilization efficiency and freedom degree for beam adjusting are higher. Compared with the optical mobile communication (OMC) system \cite{7928992,8531664}, it does not need to track users accurately in real time and can improve users' coverage by adjusting the size and distribution of sub-regions.\par
Since the core of the communication system is the array-type OIRS, we need to establish the physical models of the MA-type and OPA-type OIRS, and derive their control methods, so as to confirm the system's feasibility. In addition, the main factors affecting system performance are output power efficiency and distribution of OIRSs. The power efficiency mainly affects the large-scale fading of the system, and the power distribution mainly affects the small-scale fading introduced by pointing error. Both of them need to be deduced explicitly.
\subsection{MA-type OIRS}
The MA-type OIRS is a traditional reflecting surface with the combination of multiple mirrors, which can be implemented by micro-electro-mechanical system (MEMS) \cite{7288834,6318856,4063445}. It consists of multiple micro-mirror elements, which can freely adjust the direction of the incident beam without changing its amplitude, phase, and polarity.
\begin{figure}[htbp]
\centering
\includegraphics[width=0.95\textwidth]{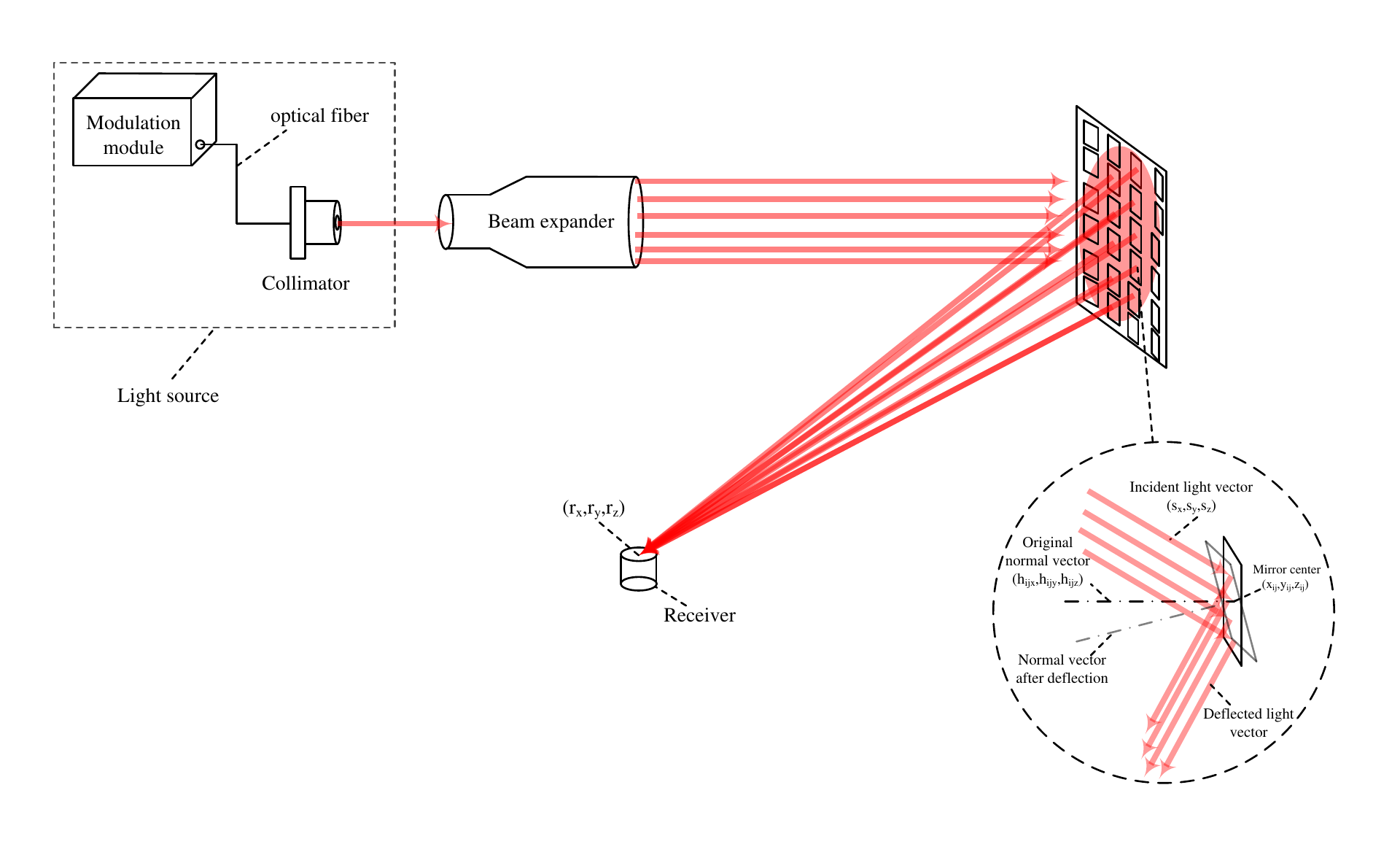}
\caption{Basic structure of MA-type OIRS-assisted optical wireless communication system.}
\label{fig.2}
\end{figure}
As can be seen from the Fig. \ref{fig.2}, when the user is at the position $(r_x,r_y,r_z)$, we need to adjust the rotation angle of each micro-mirror element to deflect the beam to the user, thereby achieving beam control. Assuming that the OIRS consists of $I\times J$ micro-mirror elements, the coordinate of the center of micro-mirror element in the i-th row and j-th column is $(x_{ij},y_{ij},z_{ij})$. The original direction vector of its normal vector is $\vec{h_{ij}}$ is $(h_{x_{ij}},h_{y_{ij}},h_{z_{ij}})$. A laser source with a collimator is utilized as light source, which can be regarded as parallel light source within a short distance under 300 meters. The beam passes through the beam expander and is expanded into a larger parallel beam to cover the OIRS surface. We assume the direction vector of the incident beam $\vec{s}$ as $(s_x,s_y,s_z)$, and the coordinates of the receiveris center as $(r_x,r_y,r_z)$. Then according to the geometric relationship, we can deduce the direction vector $\vec{h_{ij}^{'}}$ of the normal vector of the micro-mirror element in the i-th row and j-th column after deflection as
\begin{equation}
\begin{split}
\left ( \frac{r_{x}-x_{ij}}{2l_1}-\frac{s_x}{2l_2}, \frac{r_{y}-y_{ij}}{2l_1}-\frac{s_y}{2l_2},\frac{r_{z}-z_{ij}}{2l_1}-\frac{s_z}{2l_2}\right ),
\end{split}
\end{equation}
where,
\begin{equation}
\begin{split}
&l_1=\sqrt{(r_x-x_{ij})^2+(r_y-y_{ij})^2+(r_z-z_{ij})^2},\\
&l_2=\sqrt{s_x^2+s_y^2+s_z^2}.
\end{split}
\end{equation}
Therefore, the deflection angle of the micro-mirror element $\theta_{ij}$ is
\begin{equation}\label{theta}
\begin{split}
\theta_{ij}=arccos\left | \frac{h_{x_{ij}}(\frac{r_{x}-x_{ij}}{2l_1}-\frac{s_x}{2l_2})+h_{y_{ij}}(\frac{r_{y}-y_{ij}}{2l_1}-\frac{s_y}{2l_2})+h_{z_{ij}}(\frac{r_{z}-z_{ij}}{2l_1}-\frac{s_z}{2l_2})}{\sqrt{\frac{1}{2}-\frac{(r_{x}-x_{ij})s_x+(r_{y}-y_{ij})s_y+(r_{z}-z_{ij})s_z}{2l_1l_2}}} \right |
\end{split}
\end{equation}
The direction vector of the rotation axis of the normal vector $\vec{l_m}$ is
\begin{equation}
\begin{split}
\vec{l_m}&=\left (l_{m_x},l_{m_y},l_{m_z} \right )=\vec{h_{ij}}\times \vec{h_{ij}^{'}}\\&=\left (h_{y_{ij}}h^{'}_{z_{ij}}- h_{z_{ij}}h^{'}_{y_{ij}},h_{z_{ij}}h^{'}_{x_{ij}}- h_{x_{ij}}h^{'}_{z_{ij}},h_{x_{ij}}h^{'}_{y_{ij}}- h_{y_{ij}}h^{'}_{x_{ij}}\right)
\end{split}
\end{equation}
Unitize $\vec{l_m}$ to obtain $\vec{l_m^{'}}=\frac{\vec{l_m}}{\left | \vec{l_m} \right |}$.
Then we can derive the rotation matrix $R_{ij}$ of the micro-mirror element in the i-th row and j-th column as
\begin{equation}
\begin{split}
R_{ij}=Ecos\theta_{ij}+(1-cos\theta_{ij})\begin{pmatrix}
l_{m_x}
\\l_{m_y} 
\\ l_{m_z}
\end{pmatrix} \left (l_{m_x},l_{m_y},l_{m_z} \right )+sin\theta_{ij}\begin{pmatrix}
0& -l_{m_z} & l_{m_y}\\ 
l_{m_z} & 0 & -l_{m_x}\\ 
-l_{m_y} & l_{m_x} & 0
\end{pmatrix}
\end{split}
\end{equation}
According to the calculated rotation matrix group $\left[R_{11},R_{12},....R_{PQ}\right]$, we can adjust the OIRS to make each micro-mirror element reflect the beam to the target location. It can be seen from the Fig. \ref{fig.1} that in the MA-type OIRS system, the spot size at the receiver is approximately equal to the size of the micro-mirror element. Thus the size of the element need to be large enough to cover the receiver.\par

\subsection{OPA-type OIRS}
\begin{figure}[htbp]
\centering
\includegraphics[width=0.95\textwidth]{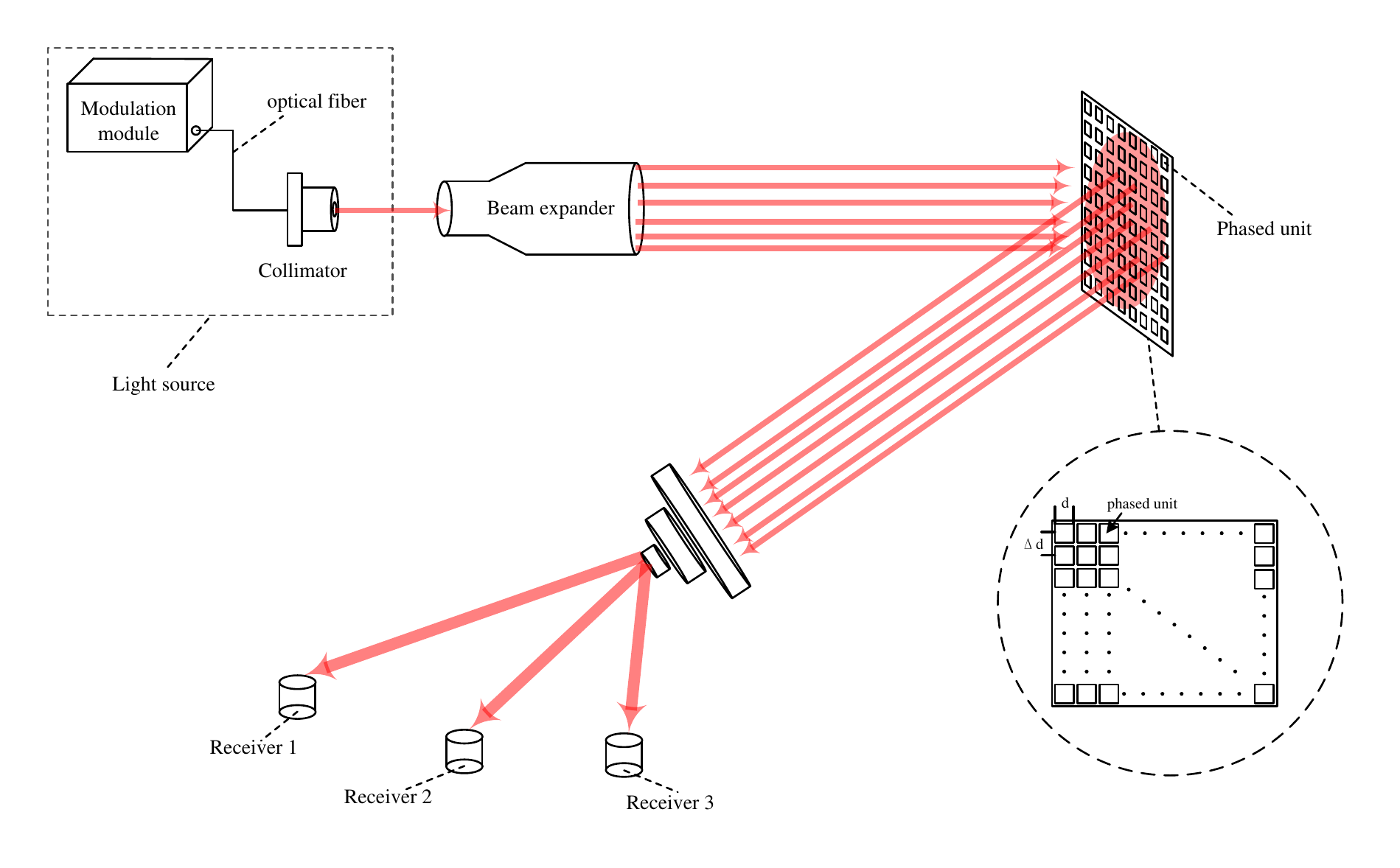}
\caption{Basic structure of OPA-type OIRS-assisted optical wireless communication system.}
\label{fig.opa1}
\end{figure}

The principle of the OPA-type OIRS is different from that of MA-type OIRS, which is mainly based on the light interference with different phases. OPA-type OIRS adjusts the optical phase distribution of the incident beam. The lens system is placed thereafter to promote the interference, thereby generating the desired light field distribution. Fig. \ref{fig.opa1} is the basic optical path diagram of OPA-type OIRS. After collimation, the laser passes through a beam expander to cover the OIRS. The beam reflected by the OIRS passes through the lens group to produce a specific interference pattern. Furthermore, the output beam is focused on the focal plane of the lens. Therefore, the receiver plane needs to coincide with the lens focal plane to achieve maximum receiving efficiency. The lens group with a variable focal length can be utilized to improve system stability. The adjustment of OPA-type OIRS is mainly based on Fraunhofer diffraction and optical holography technology. Different from MA-type OIRS system, in OPA-type OIRS system we need to determine the output light field first, and then deduce the phase distribution on OIRS. Therefore, the control algorithm of OPA-type OIRS is placed after the output light field derivation in the next section. Below we will derive the amplitude function of the light field reflected by OIRS based on its physical model.\par
As shown in Fig. \ref{fig.opa1}, we assume that there are $M\times N$ phased units in OIRS. Each unit can be regarded as a square with a side length of $d$. The spacing between adjacent phase units' center is $\Delta d$. The amplitude function of the light field reflected by OIRS $t(x,y)$ is
\begin{equation}
\begin{split}
t(x,y)=i(x,y)a(x,y)\left\{\right.rect(\frac{x}{d},\frac{y}{d})\otimes q(x,y)+\left[\right.rect(\frac{x}{\Delta d},\frac{y}{\Delta d})
-rect(\frac{x}{d},\frac{y}{d})\left.\right]\otimes p(x,y)\left.\right\},
\label{txy}
\end{split}
\end{equation}
where $i(x,y)$ is the amplitude function of incident light, which is Gaussian distribution for Gaussian source. $rect(.)$ is the rectangular function, $\otimes$ is the convolution operation, and $\emph{a(x,y)}, \emph{q(x,y)}, \emph{p(x,y)}$ are
\begin{equation}
\begin{split}
a(x,y)=rect\left(\frac{x}{M^{'}\Delta d},\frac{y}{N^{'}\Delta d}\right)\\=rect\left(\frac{x}{M^{'}\Delta d}\right)rect\left(\frac{y}{N^{'}\Delta d}\right),
\end{split}
\end{equation}
\begin{equation} \label{phase}
q(x,y)=e^{i\varphi(x,y)}\sum_{m=0}^{M^{'}-1}\sum_{n=0}^{N^{'}-1}\delta(x-m\Delta d,y-n\Delta d),
\end{equation}
\begin{equation}\label{phasec}
p(x,y)=e^{i\varphi_c(x,y)}\sum_{m=0}^{M^{'}-1}\sum_{n=0}^{N^{'}-1}\delta(x-m\Delta d,y-n\Delta d),
\end{equation}
In \eqref{phase}, $\varphi$ is the phase shift distribution on OIRS's phased units. In \eqref{phasec}, $\varphi_c$ is the phase shift superimposed on the gap between the OIRS phased units, which can not be adjusted. Due to the existence of element gap, \emph{$\Delta$d$>$d}, $\left[\right.rect(\frac{x}{\Delta d},\frac{y}{\Delta d })-rect(\frac{x}{d},\frac{y}{d})\left.\right]\otimes p(x,y)<0$. This results in the power loss of OPA-type OIRS, which is proportional to $\emph{$\Delta$d-d}$.\par
\section{Analysis of reflected beam characteristics}\label{analysis}
\subsection{Reflected beam analysis of MA-type OIRS}
\subsubsection{power density distribution}
\begin{figure}[htbp]
\centering
\includegraphics[width=0.6\textwidth]{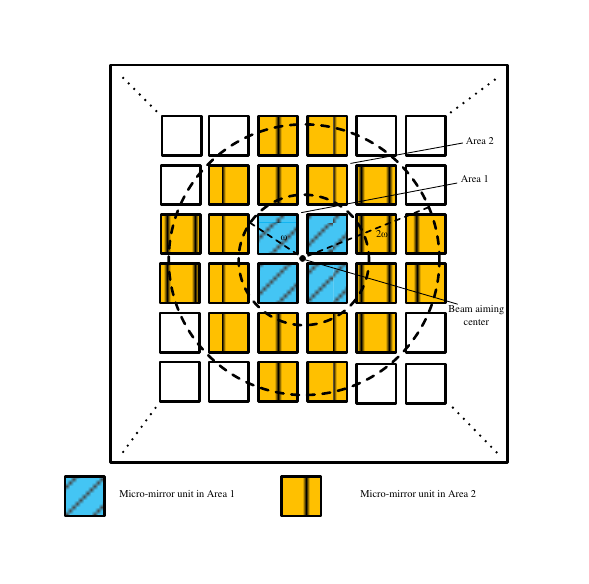}
\caption{Incident power distribution on micro-mirror elements in MA-type OIRS.}
\label{fig.3}
\end{figure}
Below we will analyze the output power distirbution from the MA-type OIRS. The beam emitted by our commonly used laser source is Gaussian light \cite{4290022,4267802,7295575}, which will pass through the beam expander after collimation in our system. The beam expander does not change the power distribution of the beam. Therefore, the beam incident on OIRS has a Gaussian distribution on its amplitude. The amplitude distribution of the Gaussian beam on the OIRS plane is 
\begin{equation}
\begin{split}
A(x,y)=\frac{A_0}{\omega_z}exp\left [ -\frac{(x^2+y^2)}{\omega_z^2} \right ]
\end{split}
\end{equation}
where $A_0$ is the central light amplitude, and $\omega_z$ is the waist radius of the beam. Since the energy of a Gaussian beam is proportional to the square of the amplitude, that is $P\propto A^2$, the beam power can be represented as  
\begin{equation}
\begin{split}
P\left ( r \right )=\kappa \frac{A_0^2}{\omega_z^2}exp\left [ -\frac{2r^{2}}{\omega_z^2} \right ]
\end{split}
\end{equation}
where $r=x^2+y^2$, and $\kappa$ is the power factor.\par
\begin{figure}[htbp]
\centering
\subfigure[Schematic diagram of the beam incident on the micro-mirror element.]{
\label{fig.4.1} 
\includegraphics[width=3.2in]{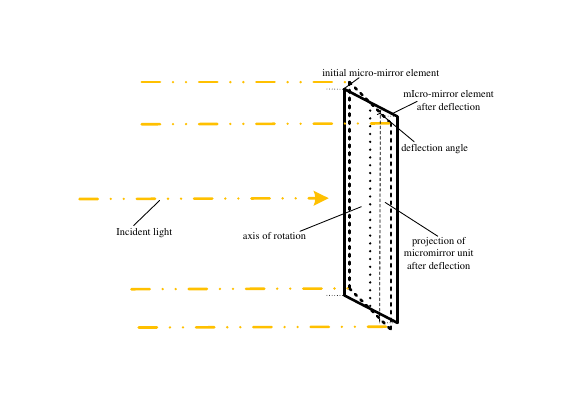}}
\hspace{1in}
\subfigure[Top view of the micro-mirror element.]{
\label{fig.4.2} 
\includegraphics[width=2in]{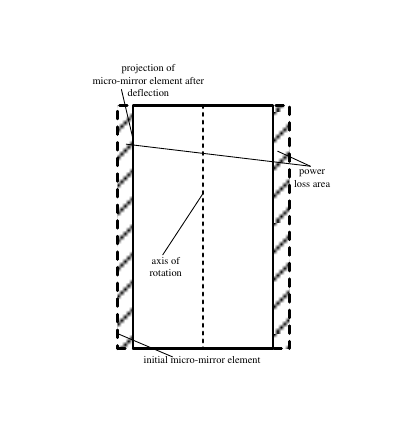}}
\caption{Schematic diagram of the micro-mirror element}
\label{fig.4} 
\end{figure}

As shown in Fig. \ref{fig.3}, the light irradiated on each micro-mirror element is a part of the Gaussian beam. When all the elements are aimed at the same target, the receiver will receive the superposition of multiple parts of the Gaussian spot, whose centers all coincide with the receiver center. For the micro-mirror element in the i-th row and j-th column whose center coordinates is $(x_{ij},y_{ij})$ in the OIRS plane, when the coordinate origin is set to its center in the receiver plane, the optical amplitude distribution from it can be represented as
\begin{equation}
\begin{split}
A_{ij}(x,y)=\left\{\begin{matrix}\frac{A_0}{\omega_z}exp\left [ -\frac{((x+x_{ij})^2+(y+y_{ij})^2)}{\omega_z^2} \right ] \quad (x,y)\in C_{ij}
\\ 
0 \quad (x,y)\notin C_{ij}
\end{matrix}\right.
\end{split}
\end{equation}
where $C_{ij}$ represents the coordinate set of the micro-mirror element in the i-th row and j-th column. The optical power density distribution from it can be represented as
\begin{equation}
\begin{split}
p_{ij}\left ( x,y \right )=\left\{\begin{matrix}\kappa \frac{A_0^2}{\omega_z^2}exp\left [ -\frac{2((x+x_{ij})^2+(y+y_{ij})^2)}{\omega_z^2} \right ] \quad (x,y)\in C_{ij}
\\ 
0 \quad (x,y)\notin C_{ij}
\end{matrix}\right.
\end{split}
\end{equation}
In addition, the deflection of the element will also bring about the loss of optical power, as shown in Fig. \ref{fig.4.1}. We set that the initial position of each micro-mirror element is perpendicular to the direction of incident light, and then the power $P_{ij}$ transmitted to the micro-mirror element in the i-th row and j-th column is 
\begin{equation}
\begin{split}
P_{ij}=\iint_{S_{ij}}p_{ij}(x,y)dxdy=\bar{P_{ij}}S_{ij},
\end{split}
\end{equation}
where $S_{ij}$ is the area of the micromirror element in the i-th row and the j-th column and $\bar{P_{ij}}$ is the average power incident on per unit area of the micro-mirror element.
Fig. \ref{fig.4.2} is a top view of the micro-mirror element. From Fig. \ref{fig.4.2}, we can observe that the deflection of the micro-mirror will bring about a power loss region, whose area is $(1-cos\theta_{ij})S_{ij}$, where $\theta_{ij}$ is the angle between the micro-mirror's deflected plane and initial plane. Therefore, we can estimate the power reflected by the deflected micro-mirror element as
\begin{equation}
\begin{split}
P_{ij}^{'}\approx \bar{P_{ij}}S_{ij}cos\theta_{ij}=P_{ij}cos\theta_{ij},
\end{split}
\end{equation}
Based on the above analysis, we can deduce that when all the elements are aimed at one target, the power density distribution from MA-type OIRS in the receiver plane $p^{'}(x,y)$ is
\begin{equation}\label{pxy}
\begin{split}
p^{'}\left ( x,y \right )=\left\{\begin{matrix}\frac{\kappa A_0^2}{\omega_z^2}\sum_{i=1}^{i=I}\sum_{j=1}^{j=J}exp\left [ -\frac{2((x+x_{ij})^2+(y+y_{ij})^2)}{\omega_z^2} \right ]cos\theta_{ij} \quad (x,y)\in C
\\ 
0 \quad (x,y)\notin C
\end{matrix}\right.
\end{split}
\end{equation}
where $C$ is the coordinate set of the OIRS output spot, whose size is similar to that of the micro-mirror element.\par
\subsubsection{power efficiency}
Although we have obtained the power distribution of OIRS output light, it is not simple for us to use it to directly calculate the output power efficiency of OIRS, since we need to determine the coordinates and size of each micro-mirror element. However, in the actual system, we can estimate the total power of OIRS by dividing the OIRS into different regions. Assume that the micro-mirror element is square and distributed as in Fig. \ref{fig.3}. Then the micro-mirror elements on OIRS can be divided into different regions according to the power distribution, as shown in Fig. \ref{fig.3}. The total power transmitted to region 1 is
\begin{equation}
\begin{split}
P_{1_{t}}=\int_{0}^{\omega}P\left ( r \right )2\pi rdr=\int_{0}^{\omega}\frac{2}{\pi\omega_z^2}P_{0}e^{-\frac{2r^2}{\omega_z^{2}}}2\pi rdr=P_{0}-P_{0}e^{-\frac{2\omega^2}{\omega_z^2}}.
\end{split}
\end{equation}
where $P_0$ is the total beam power from the transmitter, $P_0=\frac{\pi \kappa A_0^2}{2}$, and $\omega$ is the radius of region 1. There are four micro-mirror elements located in region 1, then the power transmitted to each element in region 1 can be estimated as
\begin{equation}
\begin{split}
P_{1_{e}}=\frac{1}{4}P_{0}-\frac{1}{4}P_{0}e^{-\frac{2\omega^2}{\omega_z^2}}.
\end{split}
\end{equation}
The power transmitted to each element in region k can be deduced as
\begin{equation}
\begin{split}
P_{k_{e}}=\frac{1}{4k^2}P_{0}e^{-\frac{2(k-1)^2\omega^2}{\omega_z^2}}-\frac{1}{4k^2}P_{0}e^{-\frac{2k^2\omega^2}{\omega_z^2}}.
\end{split}
\end{equation}
Considering the power loss caused by element deflection, we can deduce that when all the elements are aimed at one target, the power efficiency of MA-type OIRS $\eta_M$ is
\begin{equation}
\begin{split}
\eta_{M}=\frac{P_t}{P_0}=\sum_{k=0}^{K}\sum_{(i,j)\in C_k }(\frac{1}{4k^2}e^{-\frac{2(k-1)^2\omega^2}{\omega_z^2}}-\frac{1}{4k^2}e^{-\frac{2k^2\omega^2}{\omega_z^2}})cos\theta _{ij},
\end{split}
\end{equation}
where $P_t$ represents total power reflected by OIRS, $C_k$ represents the set of micro-mirror elements located in the region k, and $K$ represents the number of regions on OIRS.
The result of this estimation is highly related to the distribution of micro-mirror elements on the OIRS and the position of the beam center. In a practical system, the relative position of the OIRS and transmitter is always fixed. Once the distribution of the micro-mirror elements on OIRS and the position of beam center (usually pointing to the center of the OIRS) are fixed, we can estimate the power distribution and efficiency by this way. Therefore, although the derivation result is based on a specific element distribution on OIRS, the deduction method does not deviate from the actual system.\par
\subsection{Reflected beam analysis of OPA-type OIRS}\label{OPA}
\subsubsection{Fraunhofer diffraction}
\begin{figure}[htbp]
\centering
\includegraphics[width=0.7\textwidth]{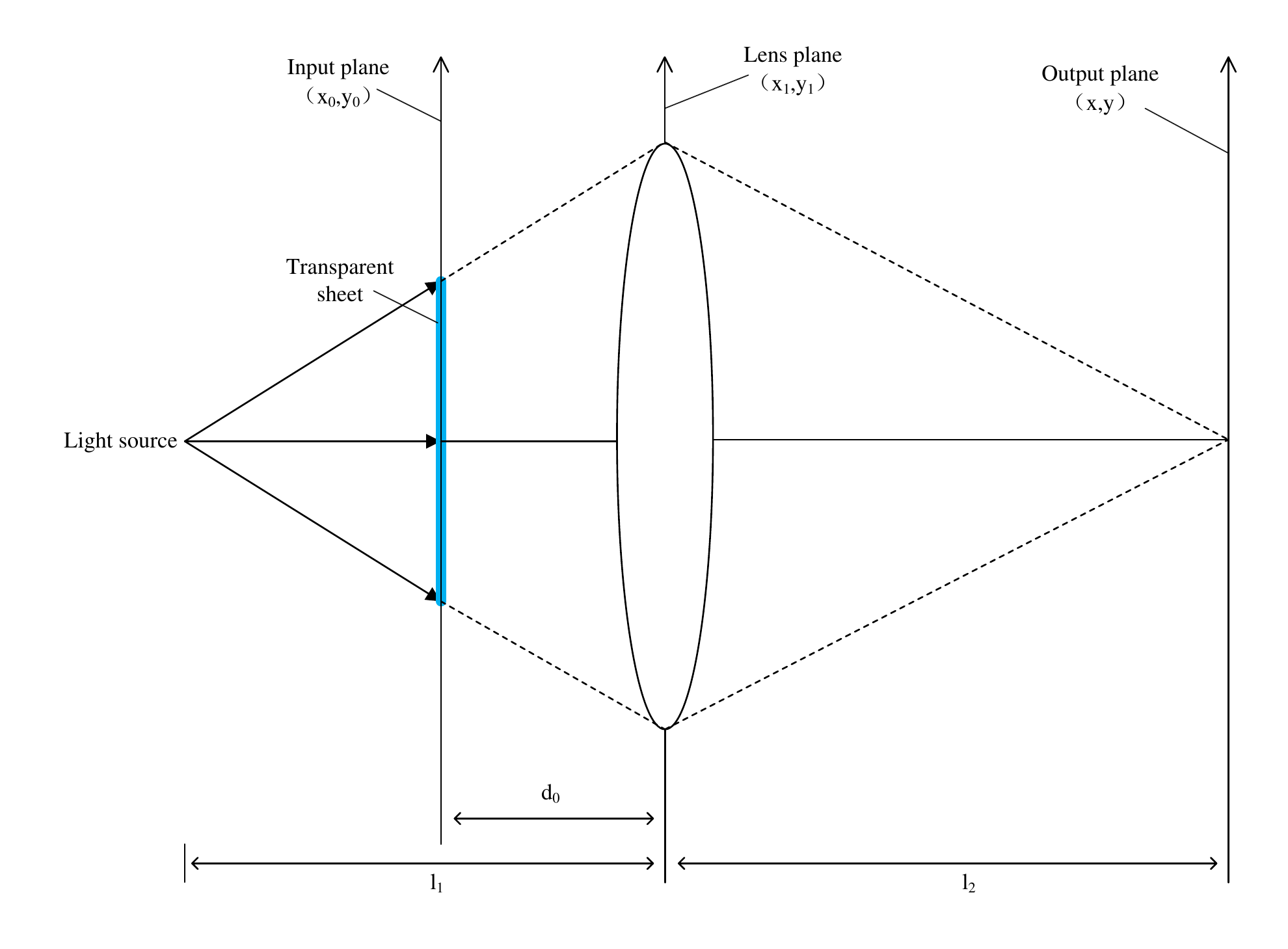}
\caption{Diagram of the Fraunhofer diffraction.}
\label{fig.mirror}
\end{figure}
The beam reflected by OIRS will undergo Fraunhofer diffraction through the lens \cite{8734708,5459086}. As shown in Fig. \ref{fig.mirror}, we place a transparent sheet in front of the light source as the input surface, and set its complex amplitude transmittance to $r(x_0,y_0)$. Then, the complex amplitude of the light field formed by the spherical wave from light source is $R_{0}e^{jk\frac{x_{0}^{2}+y_{0}^{2}}{2(l_{1}-d_{0})}}$, where $R_{0}$ is the amplitude of the light source, $k$ is the wave number, $k=\frac{2\pi}{\lambda}$, and $\lambda$ is the light wavelength. The complex amplitude function of the light field emitted from the input plane $U_0$ is
\begin{equation}
\begin{split}
U_0=R_0r(x_0,y_0)e^{jk\frac{x_0^2+y_0^2}{2(l_1-d_0)}}
\end{split}
\end{equation}
When light reaches the lens plane from the input plane, Fresnel diffraction occurs in space. According to the Fresnel diffraction formula, the complex amplitude of the light field in front of the lens plane $U_1$ is
\begin{equation}
\begin{split}
U_1(x_1,y_1)=\frac{R_0}{j\lambda d_0}\iint r(x_0,y_0)e^{jk\frac{x_{0}^{2}+y_{0}^{2}}{2(l_1-d_0)}}e^{jk\frac{(x_1-x_0)^2+(y_1-y_0)^2}{2d_0}}dx_0dy_0
\end{split}
\end{equation}
The light field distribution after passing through the lens $U_2$ is
\begin{equation}
\begin{split}
U_2(x_1,y_1)=U_1(x_1,y_1)P(x_1,y_1)e^{-jk\frac{x_1^2+y_1^2}{2f}}
\end{split}
\end{equation}
where $P(x_1,y_1)$ is the pupil function, and $f$ is the focal length of the lens. Since the size of the lens is not infinite, the light outside the lens will not be imaged, that is, inside the lens, $P\left (x_1,y_1 \right )=1$, outside the lens, $P\left (x_1,y_1 \right) =0$. The light field after passing through the lens will undergo Fresnel diffraction again and reach the output plane. The light field distribution function $U_3$ of the output plane is
\begin{equation}
\begin{split}
U_3(x,y)=\frac{1}{j\lambda l_2}\iint U_2(x_1,y_1)e^{jk\frac{x_{1}^{2}+y_{1}^{2}}{2f}}e^{jk\frac{(x-x_1)^{2}+(y-y_1)^{2}}{2l_2}}dx_1dy_1
\end{split}
\end{equation}
Substituting $U_2$ we can obtain that
\begin{equation}\label{U3}
\begin{split}
U_3(x,y)=Be^{jk\frac{(f-d_0)(x^2+y^2)}{2l_2(f-d_0)+2fd_0}}\iint_{-\infty}^{\infty} r(x_0,y_0)e^{-jk\frac{(x_0x+y_0y)f}{l_2(f-d_0)+fd_0}}dx_0dy_0
\end{split}
\end{equation}
In \eqref{U3}, $B$ is a constant. We can observe that $\iint_{-\infty}^{\infty} r(x_0,y_0)e^{-jk\frac{(x_0x+y_0y)f}{l_2(f-d_0)+fd_0}}dx_0dy_0 $ is the Fourier transform of the transmittance function of the input surface, that is, the Fourier transform of the original object light wave function formed by the object light wave through the lens. Therefore, to restore the original light wave function distribution of the output surface light field function, the input surface should be loaded with the inverse Fourier transform of the object light wave light field.\par
\subsubsection{power distribution and efficiency} \label{pd}
According to the above analysis, the optical field amplitude distribution function $T(u,v)$ after Fraunhofer diffraction is the Fourier transform of the OIRS reflectivity function $t(x,y)$. That is, $T(u, v)=F\left \{ t(x,y) \right \}$, where $F\left \{ \right \}$ is the Fourier transform. Substituting into \eqref{txy}, we can obtain
\begin{equation}\label{Tuv}
\begin{split}
T\left ( u,v \right )=I\left ( u,v \right )\otimes A\left ( u,v \right )\otimes \left \{ d^{2}sinc(ud,vd)Q(u,v)+\left [ \Delta d^{2}sinc(u\Delta d,v\Delta d)-d^{2}sinc(ud,vd) \right ]P(u,v) \right \},
\end{split}
\end{equation}
where
\begin{equation}
\begin{split}
&I\left ( u,v \right )=F\left \{ i(x,y) \right \},\\
&A\left ( u,v \right )=F\left \{ a(x,y) \right \},\\
&sinc(u,v)=\frac{sin(\pi u)}{\pi u}\frac{sin(\pi v)}{\pi v},\\
&Q(u,v)=F\left \{ exp(i\varphi ) \right \}\otimes\sum_{m,n=-\infty }^{\infty }\delta \left ( u-\frac{m}{\Delta d},v-\frac{n}{\Delta d} \right ),\\
&P(u,v)=F\left \{ exp(i\varphi_c )\right \}\otimes\sum_{m,n=-\infty }^{\infty }\delta \left ( u-\frac{m}{\Delta d},v-\frac{n}{\Delta d} \right ).
\end{split}
\end{equation}
We can observe that $I\left ( u,v \right )\otimes A\left ( u,v \right )\otimes \left [ \Delta d^{2}sinc(u\Delta d,v\Delta d)-d^{2}sinc(ud,vd) \right ]P(u,v)$ in \eqref{Tuv} is not adjustable, since $\varphi_c(x,y)$ is the phase distribution of gap between phased units. Regarding it as power loss, we can obtain the output power efficiency of OPA-type OIRS $\eta _{O}$ as
\begin{equation}
\begin{split}
\eta _{O}=\frac{\iint_{u,v=-\infty }^{u,v=\infty}\left [ I\left ( u,v \right )\otimes A\left ( u,v \right )\otimes d^{2}sinc(ud,vd)Q(u,v)\right ]^2dudv}{\iint_{u,v=-\infty }^{u,v=\infty}T^2(u,v)dudv}.
\end{split}
\end{equation}
From the above derivation, we can observe that the input phase distribution on the OPA-type OIRS is determined by the output light field distribution. Therefore, in addition to the non-adjustable light field brought by the phased units' gap, the OIRS output light field distribution can be determined according to the needs of the system. For example, we can set the output optical power to a uniform distribution, then the impact of pointing error on system performance will be greatly reduced.
\subsubsection{derivation of phase distribution on OIRS}
According to the above derivation, when we determine the expected light field distribution $E(u,v)$, we can inversely deduce the phase distribution on OIRS according to \eqref{txy} and \eqref{Tuv}. However, if we directly substitute $E(u,v)$ as $T(u,v)$ into \eqref{Tuv}, the non-adjustable light field introduced by the unit gap may cause the tunable light field distribution to appear negative. Therefore, we need to superimpose $E(u,v)$ on the non-adjustable light field to obtain $T(u,v)$, namely
\begin{equation} \label{Tuv2}
\begin{split}
T(u,v)=E(u,v)+I\left ( u,v \right )\otimes A\left ( u,v \right )\otimes \left [ \Delta d^{2}sinc(u\Delta d,v\Delta d)-d^{2}sinc(ud,vd) \right ]P(u,v).
\end{split}
\end{equation}
Substituting into \eqref{txy} and \eqref{Tuv}, we can obtain the phase distribution of OIRS units $\varphi(x,y)$. In the actual system, we can import the phase distribution to the OIRS to output the desired light field.
\section{Beam splitting and power allocation}\label{split}
\subsection{Beam splitting for MA-type OIRS}
MA-type OIRS can realize the functions of beam splitting and power allocation in point-to-multipoint FSO scenarios. According to the above analysis, we can adjust the beam convergence by adjusting the deflection direction of micro-mirror elements to focus the beam on certain targeting points. At the same time, we can also group the micro-mirror elements, and focus the beam on multiple users by selecting different groups of micro-mirror elements for deflection. Therefore, on the basis of the above mathematical model, this problem can be equivalent to an optimization problem of grouping micro-mirror elements.\par
Assume that the beam is divided into $m$ sub-beams to cover $m$ sub-areas through the OIRS. The power ratio of the sub-beams is $k_1:k_2:...:k_m$. According to the analysis of Section \ref{model}, when the sub-area is determined, the deflection angle of the OIRS micro-mirror element is uniquely determined, which corresponds to a specific output power. Therefore, we can establish the mapping between sub-area $n$ and reflecting power matrix $R_n$, that is
\begin{equation}
\begin{split}
n\rightarrow R_n=\begin{bmatrix}
p_{11}cos\theta_{11}^{(k)} & p_{12}cos\theta_{12}^{(k)}  & \cdots  & \cdots & p_{1N}cos\theta_{1N}^{(k)}\\ 
p_{21}cos\theta_{21}^{(k)}& \ddots  & & & p_{2N}cos\theta_{2N}^{(k)}\\ 
\vdots  &  & p_{ij}cos\theta_{ij}^{(k)} & &\vdots\\ 
\vdots& & & \ddots &\vdots\\
p_{M1}cos\theta_{M1}^{(k)} & p_{M2}cos\theta_{M2}^{(k)} & \cdots & \cdots&p_{MN}cos\theta_{MN}^{(k)},
\end{bmatrix}
\end{split}
\end{equation}
where $p_{ij}$ is the total light power incident on the micro-mirror element in the i-th row and j-th column, which is irrelevant to the deflection of the micro-mirror element, $\theta_{ij}$ is the deflection angle of the micro-mirror element in the i-th row and j-th column. The optimization problem can be expressed as
\begin{equation}\label{optimal}
\begin{split}
\left\{\begin{matrix}
&P_k=\sum_{i,j\in M_k}p_{ij}cos\theta_{ij}^{(k)}
\\ s.t. \quad &P_1:P_2:\cdots:P_m=k_1:k_2:\cdots:k_m
\\ &M_i\bigcap M_j=\varnothing \quad (i\neq j,1\leq i,j\leq m)
\\ &M_i\subseteq R \quad (1\leq i\leq m )\\
max &\sum_{k=1}^{m}P_k
\end{matrix}\right.,
\end{split}
\end{equation}
where $M_k$ represents the set of micro-mirror elements that reflect the beam to sub-area $k$, $R$ represents the set of micro-mirror elements on OIRS. The solution of the optimization problem represents that the system selects the optimal micro-mirror elements' group corresponding to each subregion. Therefore, the OIRS output power is maximized under the premise of ensuring that the sub-beam power ratio conforms to the system configuration.
\subsection{Beam splitting for OPA-type OIRS}
According to the derivation in Section \ref{OPA}, to achieve beam splitting and power allocation, we need to determine the expected output light field and deduce the phase distribution of OIRS unit. Assume that the light beam is divided into $m$ sub-beams by OIRS reflection to cover $m$ sub-regions. The power ratio of the sub-beams is $k_1:k_2:\cdots:k_m$. We can set the expected output light field as
\begin{equation}
\begin{split}
E(u,v)=\sum_{i=1}^{m}A_{i}E_{i}(u,v).
\end{split}
\end{equation}
where $A_i$ the amplitude in subregion i, $E_{i}(u,v)$ is the spatial distribution of subregion i and 
\begin{equation}
\begin{split}
&A_1:A_2:\cdots:A_m=\sqrt{k_1}:\sqrt{k_2}:\cdots:\sqrt{k_m},\\
&E_i(u,v)\cap E_j(u,v)=\varnothing,  \quad i\neq j, 1 \leq i,j\leq m.
\end{split}
\end{equation}
The light field distribution describes $m$ non-coincident sub-regions, and the light amplitude in the region is uniformly distributed. Substituting it into \eqref{Tuv2}, and \eqref{Tuv}, the OIRS phase distribution can be obtained. In addition, we can regard the  light field $I\left ( u,v \right )\otimes A\left ( u,v \right )\otimes \left [ \Delta d^{2}sinc(u\Delta d,v\Delta d)-d^{2}sinc(ud,vd) \right ]P(u,v)$ as a sub-area. Since the phase shift of the unit gap is approximately constant, the non-adjustable light field energy will be concentrated at the focal point of the lens. This part of the power can be utilized as a fixed sub-beam to increase power efficiency.
\section{Discussion}\label{discussion}
\subsection{Channel fading caused by pointing error for systems with two types of OIRSs}
In FSO systems, the small-scale channel fading that affects system performance mainly comes from pointing errors and atmospheric turbulence \cite{6844864,1025501,4267802}. Since OIRS does not change the characteristics of the atmospheric channel, the channel fading caused by atmospheric turbulence in OIRS-assisted FSO system performs the same as that in traditional communication system \cite{9013840,Naja2019,Naja2020}. Pointing error mainly comes from the jitter of the beam at the transmitting end, which causes the center of the receiver to deviate from the center of the Gaussian beam \cite{borah2009pointing,sandalidis2008ber,burks1982high,trung2014pointing,borah2009pointing}. According to this characteristic, we can deduce the performance loss of the system. However, when the signal is not Gaussian distributed, the performance loss caused by the random offset of the receiver center position will be completely different. For example, assuming that the signal light is uniformly distributed and completely covers the receiver, the offset of the receiver center will hardly bring in performance loss.\par
For MA-type OIRS, the output optical power density is distributed as shown in \eqref{pxy}, which indicates that the peak output power may not be located in the center of the receiver. Therefore, specific values of pointing error may even bring performance gains. Although the output light field of OPA-type OIRS can be uniformly distributed, pointing error will still bring in a certain performance impact. For OPA-type OIRS, the initial input light field set in \eqref{txy} is Gaussian, which does not include the offset caused by pointing error. Therefore, when we use the phase distribution derived from \eqref{txy} to control the OIRS, since the input light field containing pointing error is superimposed on OIRS, the output light field will not be uniformly distributed.
\subsection{Comparison of two types of OIRSs in the actual system}
In this article, we deduced the control methods, output power distribution, power efficiency, algorithms of beam splitting and power allocation for two types of OIRSs. Although these characteristics are highly related to the number, size and distribution of OIRS units, we can still make a comparison between two types of OIRS in certain aspects.\par
In terms of the freedom of beam control, the OPA-type OIRS can achieve more arbitrary light intensity distribution on the lens focal plane. Since the size of the MA-type OIRS output beam is approximately the same as that of the micro-mirror element, it is difficult to adjust the sub-rgion's size for MA-type OIRS. To cover a larger region, the MA-type OIRS system need to spread out multiple spots from different micro-mirror elements, which is not as good as the OPA-type OIRS in terms of control flexibility and output beam quality. \par
In terms of the complexity of the control algorithm, the control complexity of a single unit of MA-type OIRS is greater than that of OPA-type OIRS. However, since the number of OPA-type OIRS units per unit area is much larger than that of MA-type OIRS, it is difficult to directly compare the control algorithm complexity of the two types of OIRS. In the experiment of this article, we measured the operation time of the MA-type control algorithm is $0.45 ms$, and that of the OPA-type control algorithm is $0.36 ms$ for beam splitting of three sub-regions.\par
In terms of output power efficiency, the power loss of MA-type OIRS mainly comes from the micro-mirror gap caused by its deflection, which is related to the target's position and difficult to eliminate. The output power loss of OPA-type OIRS mainly comes from the gap between phased units. Under the existing technology, the OPA's fill factor (the ratio of the area of the phased units to the area of the OPA) can reach $98\%$, which is expected to be further improved in the future. Therefore, the output power efficiency of OPA-type OIRS is theoretically higher than MA-type OIRS. In our experiment, we measured the output power efficiency of two types of OIRSs for beam splitting of three sub-regions, where the power efficiency of MA-type OIRS is $89.72\%$, and that of OPA-type OIRS is $95.64\%$.\par
In terms of channel fading and communication performance, according to \eqref{pxy}, \eqref{Tuv}, Figs. \ref{fig.mems} and \ref{fig.opa} and the discussion of pointing error, we can observe that the distribution of MA-type OIRS output power density has a large spatial variation trend. Therefore, the deviation of the beam center caused by pointing error has a greater impact on system performance. OPA-type OIRS output power distribution is more even and smooth, which is beneficial to the performance of the communication system.\par
In terms of cost, MA-type OIRS requires fewer units with lower cost. Due to the need to fit the light wavelength, the phased units' size of OPA-type OIRS is only micron level. Therefore, it is difficult to occupy a large area with an OPA-type OIRS. The utilization of OPA-type OIRS will also increase the accuracy requirements for the base station's targeting.
\section{Numerical Results}\label{simulation}
\subsection{Experimental setup}
\begin{figure}[htbp]
\centering
\includegraphics[width=0.7\textwidth]{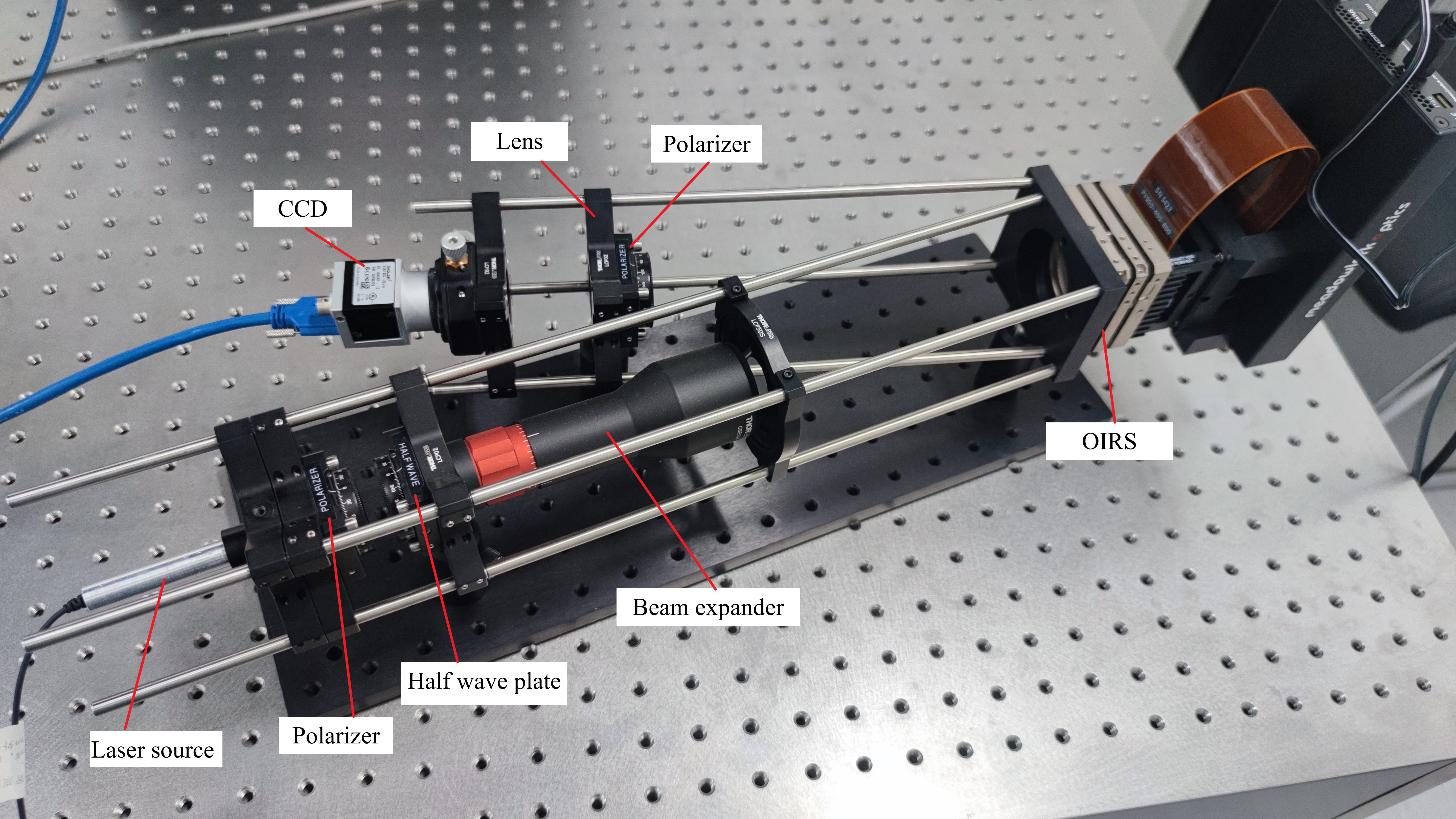}
\caption{Experimental light path.}
\label{fig.exp}
\end{figure}

We built an experimental device to verify the accuracy of the model. The experimental optical path is the same as that shown in Fig. \ref{fig.2} and \ref{fig.opa1}, where the receiving end is replaced by a charge coupled device (CCD) to observe the output pattern or an optical power probe to detect the power distribution. The actual experimental light path is shown in Fig. \ref{fig.exp}. We utilize a 532nm laser source in the system. After collimated by the collimator, the parallel-like light with extremely small emission angle is obtained. The light beam reaches the beam expander after passing through the polarizer and half-wave plate, and is expanded into a larger parallel beam to cover the OIRS. In the MA-type OIRS system, the beam reflected by the OIRS is directly received by the CCD or optical power sensor. In the OPA-type OIRS system, the light beam needs to pass through a convex lens with focal length of $25 cm$. The CCD or optical power sensor is placed at the focal length of the lens to obtain the highest power receiving efficiency. In addition, we utilize $16$ MEMS scanning mirrors (diameter of $40 mm$, deflection angle resolution of $10 \mu rad$) to form a $4\times 4$ micro-mirror array as the MA-type OIRS. A pure phased modulated liquid crystal optical phase control array ($1920 \times 1152$ pixels, fill factor $95.7\%$, size $17.7 \times 10.6 mm$) is adopted as the OPA-type OIRS. Due to the limitation of the experimental site and CCD area, we utilize the MEMS, OPA and lens with centimeter size in this experiment. In the actual system, we can increase the coverage area of the beam by using a large-area micro-mirror, OPA and lens group.
\subsection{Optical power distribution}
\begin{figure}[htbp]
\centering
\includegraphics[width=0.7\textwidth]{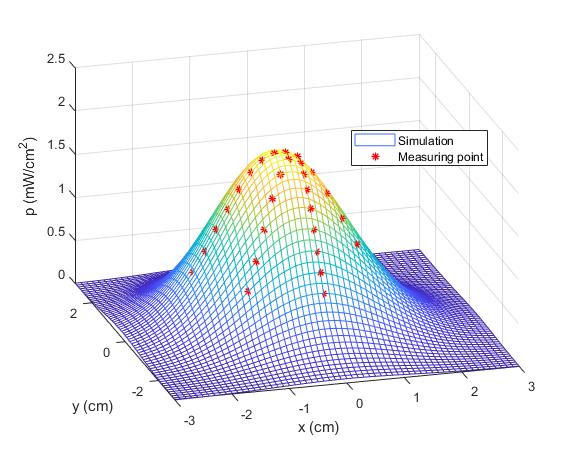}
\caption{Power density distribution from MA-type OIRS.}
\label{fig.mems}
\end{figure}

\begin{figure}[htbp]
\centering
\includegraphics[width=0.7\textwidth]{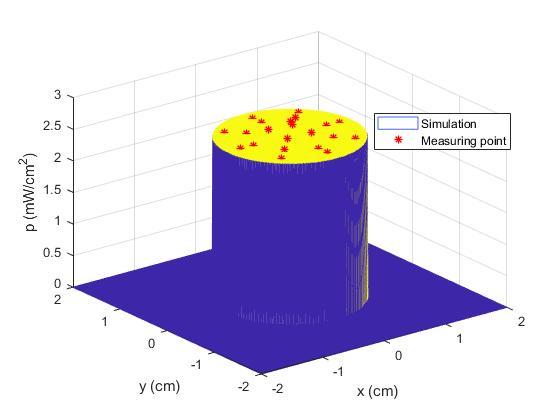}
\caption{Power density distribution from OPA-type OIRS.}
\label{fig.opa}
\end{figure}

First, we need to experimentally verify the output power distribution models of the two types of OIRS. A two-dimensional rectangular coordinate system is set up with the directions of the two rectangular sides of OIRS as the $x$ and $y$ axes. The origin is set on a straight line perpendicular to the OIRS plane and passing through the center of the OIRS, $25 cm$ from the center of the OIRS. According to \eqref{pxy} and \eqref{Tuv2}, we simulated the optical power distribution of the two types of OIRS on the receiving plane when the desired optical field center is set at the coordinate $(0, 0)$. The size of the MA-type OIRS element is set the same as that of the experimental device, and the deflection angle of each micro-mirror element is obtained according to \eqref{theta}.The expected output light field of OPA-type OIRS is set to uniform distribution, and the output light spot diameter is set to $2cm$. Its phase distribution is derived from \eqref{Tuv}. At the same time, we measured the power at multiple coordinate positions (To improve the measurement accuracy, we choose an optical power probe with $2mm$ diameter. The measured power is divided by the probe area to estimate the power density at this point) and compared them with the simulation results, as shown in Figs. \ref{fig.mems} and \ref{fig.opa}. The measured power is basically consistent with the simulation results, which proves the power distribution derivation in this paper is accurate. For OPA-type OIRS, we observe that the measured power near the origin is not consistent with the expected power. This results from the non-adjustable light field that we mentioned in Section \ref{pd}, whose power focuses on the focal point of the lens. For MA-type OIRS, we can observe that the output power density distribution is different from that of Gaussian light, whose power peak is not at the origin and the gradient varies along different directions.
\subsection{Beam splitting and power allocation}
\begin{figure}[htbp]
\centering
\includegraphics[width=0.7\textwidth]{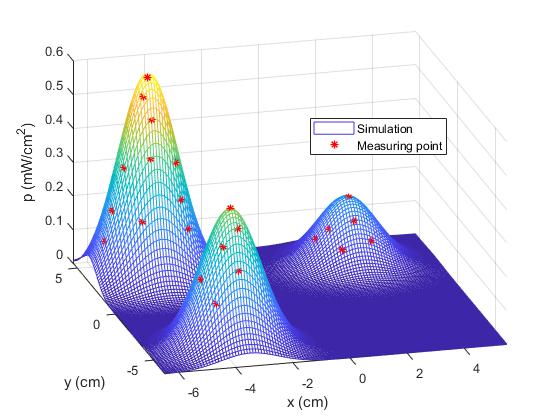}
\caption{Power density distribution for beam splitting from MA-type OIRS.}
\label{fig.mems2}
\end{figure}

\begin{figure}[htbp]
\centering
\includegraphics[width=0.7\textwidth]{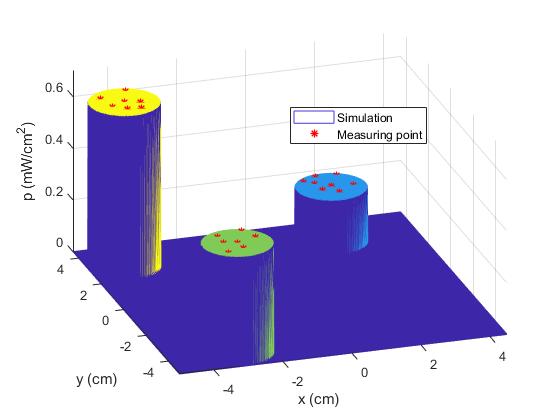}
\caption{Power density distribution for beam splitting from OPA-type OIRS.}
\label{fig.opa2}
\end{figure}

We adopt two types of OIRSs for beam splitting to cover multiple sub-areas. The center coordinates of the sub-area are respectively $(2,3),(-3,-4),(-4,3)$, which corresponds to a power ratio of 1:2:3. In MA-type OIRS system, we use \eqref{optimal} to calculate the optimal allocation of mico-mirror elements. In the OPA-type OIRS system, we set the expected light field of each sub-area to be uniformly distributed with a diameter of $2cm$, and set a small block at the focal point of the lens to filter out the influence of the non-adjustable light field. At the same time, we select multiple coordinate positions for optical power measurement and compared them with the simulation results, as shown in Figs. \ref{fig.mems2} and \ref{fig.opa2}. The measured power is basically consistent with the simulation result, which proves that the beam splitting model control algorithm in this paper is accurate. We also observe that the power ratio of the two types of OIRS sub-regions roughly matches the expected power ratio, which proves that the two types of OIRS can achieve accurate power distribution. 

\section{Conclusion}\label{conclusion}
In this work, we derive the control algorithm, beam splitting algorithm, output power density distribution and output power efficiency of MA and OPA-type OIRSs. Experiments have proved that our model and deduced results fit the reality and the control algorithm is feasible. According to the results of this paper, OPA-type OIRS has certain advantages over MA-type OIRS in terms of freedom of output beam control, output power efficiency and communication performance under small-scale fading. However, the MA-type OIRS requires lower number of units and cost per unit area. In addition, the power density distribution of two types of OIRSs derived in this paper can become the basis for performance analysis of communication system assisted by array-type OIRSs. In the subsequent research, we will solve the system's channel fading and performance, and further improve the related theories of array-type OIRSs and point-to-multipoint distributed FSO systems.


%

\ifCLASSOPTIONcaptionsoff
  \newpage
\fi



%

%

\bibliographystyle{IEEEtran}
\bibliography{IEEEabrv,ref}
\end{document}